\documentclass[a4paper,11pt]{article}
\usepackage{pos}

\usepackage{amsmath,amssymb}
\usepackage{slashed}
\usepackage{graphicx}

\title{Jet functions for next-to-leading power factorization}
\ShortTitle{Jet functions at NLP}

\author[a]{Robin van Bijleveld}
\author[b]{Jaco ter Hoeve}
\author[a,c,d]{Eric Laenen} 
\author[a]{Coenraad Marinissen}
\author*[e]{Leonardo Vernazza}
\author[f]{Guoxing Wang}

\affiliation[a]{Nikhef, Theory Group, Science Park 105, 1098 XG, Amsterdam, The Netherlands}
\affiliation[b]{The Higgs Centre for Theoretical Physics, University of Edinburgh,
JCMB, KB, Mayfield Rd, Edinburgh EH9 3FD, Scotland}
\affiliation[c]{IoP/ITFA, University of Amsterdam, Science Park 904, 1098 XH Amsterdam, The Netherlands}
\affiliation[d]{ITF, Utrecht University, Leuvenlaan 4, 3584 CE Utrecht, The Netherlands}
\affiliation[e]{INFN, Sezione di Torino, Via P. Giuria 1, I-10125 Torino, Italy}
\affiliation[f]{Laboratoire de Physique Th\'eorique et Hautes Energies (LPTHE), UMR 7589, Sorbonne Universit\'e et CNRS, 4 place Jussieu, 75252 Paris Cedex 05, France}

\emailAdd{leonardo.vernazza@to.infn.it}

\abstract{%
We discuss the factorization of scattering processes near
partonic threshold at next-to-leading power (NLP) in the
threshold variable $1-z$, with $z \equiv q^2/\hat{s}$. We
review the general structure of power-suppressed contributions
both in Soft-Collinear Effective Theory (SCET) and in a direct
QCD approach, and discuss the definition of NLP jet functions
as gauge-invariant operator matrix elements in QCD. As a
controlled check of the resulting factorization formula, we
verify it explicitly at one and two loops for the massive
electromagnetic form factor in the limit $m^2 \ll s$, using the
method of regions. We conclude by outlining the two challenges
that remain for a systematic resummation of NLP logarithms: the
treatment of endpoint divergences in SCET convolutions, and the
extension of jet functions to radiative processes capable of
describing an arbitrary number of soft-gluon emissions -- the
latter being the last missing ingredient for exponentiation,
given that the purely soft sector is already understood in
terms of generalised webs via the replica trick.
}

\FullConference{Loops and Legs in Quantum Field Theory (LL2026)\\
12--17 April, 2026\\
Bayreuth, Germany\\}

\begin{document}
\maketitle

\section{Introduction: particle scattering near threshold}

A key challenge in precision phenomenology
at hadron colliders is the resummation of
large logarithms beyond leading power.
As an example, consider Drell-Yan production
$q(p_1)\bar{q}(p_2)\to\gamma^*(q)\to l^+ l^-$,
with $\hat{s}\equiv(p_1+p_2)^2$ the partonic
centre-of-mass energy, and the threshold
variable $z\equiv q^2/\hat{s}\to 1$.
The partonic cross section admits a
singular expansion in $(1-z)$:
\begin{equation}
\Delta_{ab}(z) \, \sim \sum_{n=0}^{\infty} \left(\frac{\alpha_s}{\pi}\right)^n
\!\left[ c_n\,\delta(1-z)
  + \sum_{m=0}^{2n-1} c_{nm}
    \left[\frac{\ln^m(1-z)}{1-z}\right]_{\!\!+}
  + d_{nm}\ln^m(1-z)
  + \ldots \right].
\end{equation}
The $\delta(1-z)$ and plus-distribution
terms are the leading-power (LP) contributions;
the regular logarithms $\ln^m(1-z)$ are suppressed
by one power of $(1-z)$ and define the next-to-leading
power (NLP) terms.

\begin{figure}[h]
\begin{center}
  \includegraphics[width=0.50\textwidth]{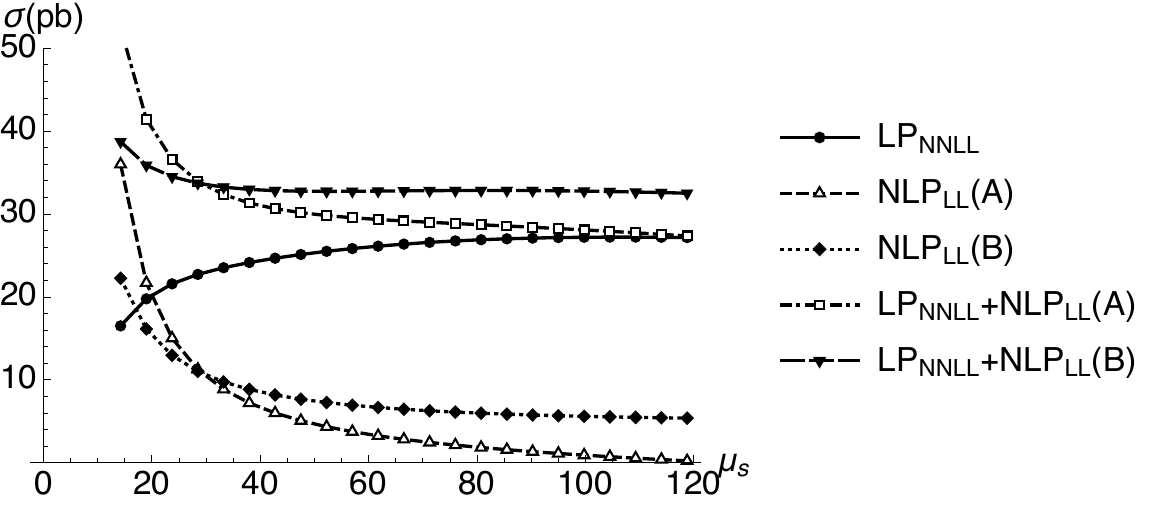}
\end{center}
\caption{LP vs.\ NLP resummation in SCET at
various logarithmic accuracies for single
Higgs production, from \cite{Beneke:2019mua}.}
  \label{Pheno-NLP-1}
\end{figure}
\begin{figure}[h]
\begin{center}
  \includegraphics[width=0.40\textwidth]{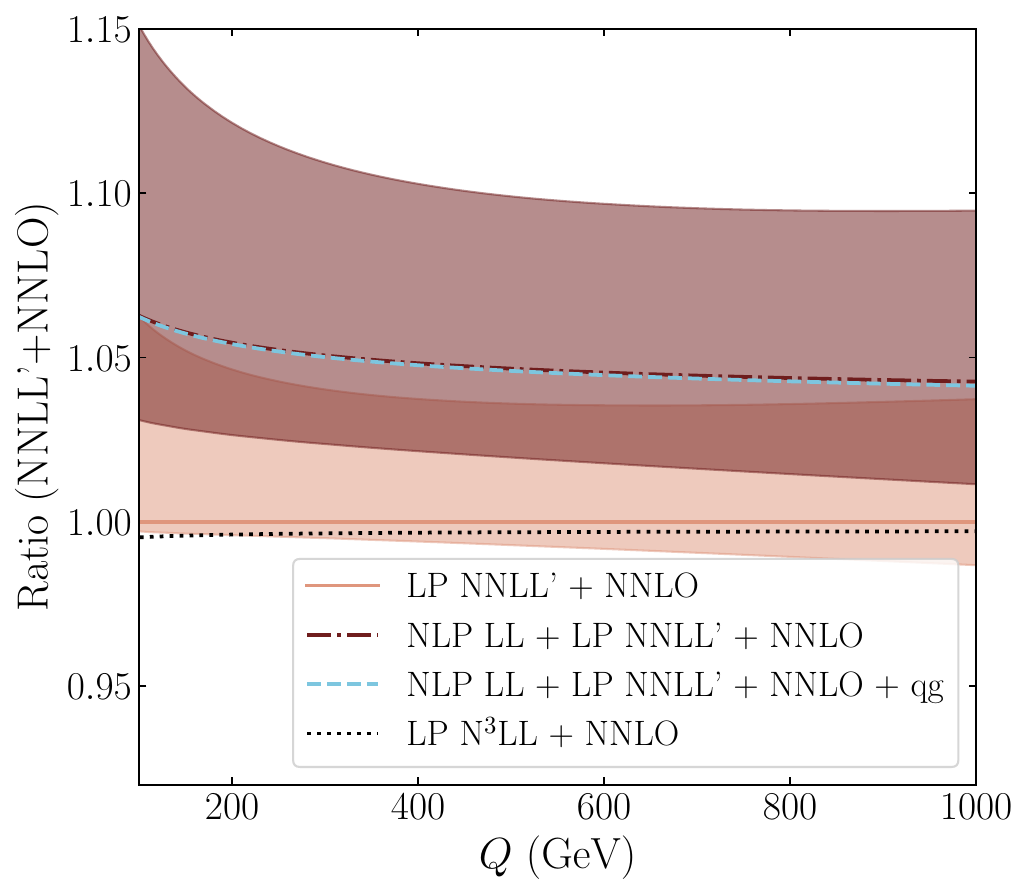}
  \includegraphics[width=0.40\textwidth]{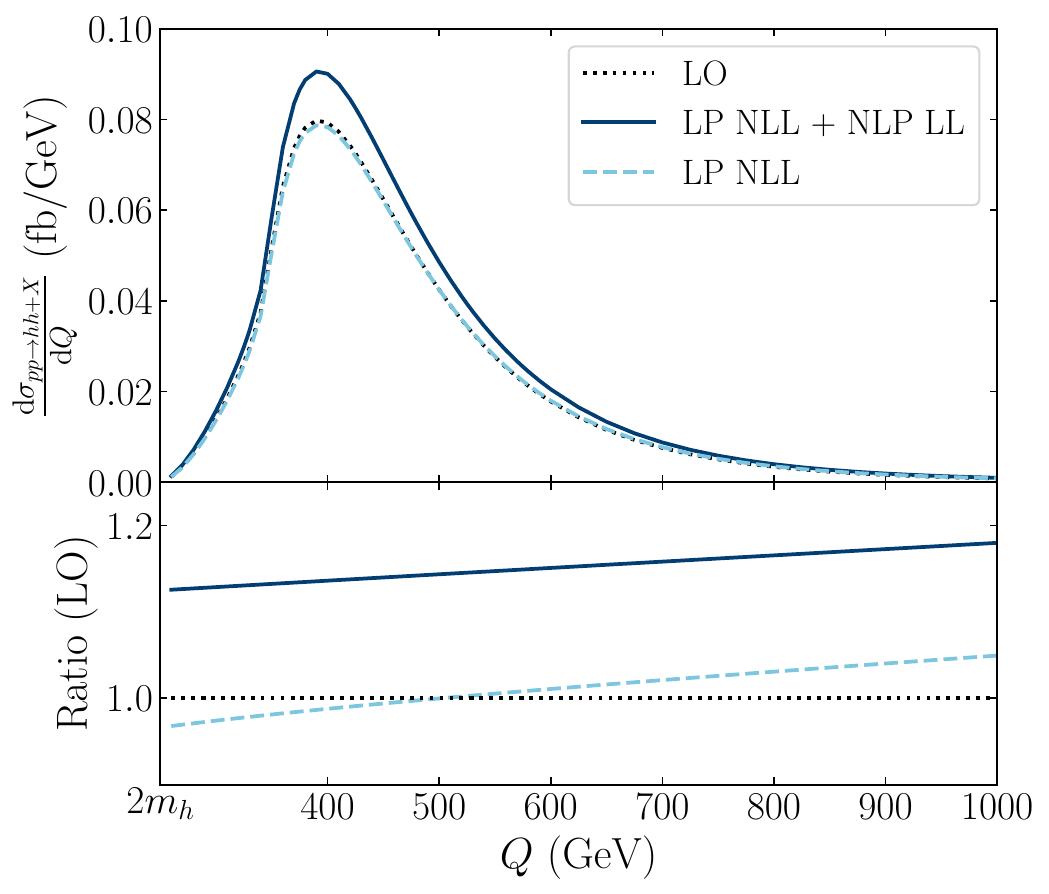}
\end{center}
\caption{LP vs.\ NLP resummation in QCD at
various logarithmic accuracies for single
(left) and double (right) Higgs production,
from \cite{vanBeekveld:2021hhv}.}
  \label{Pheno-NLP-2}
\end{figure}

The resummation of LP logarithms is
well understood up to N$^3$LL and beyond; see
\cite{Sterman:1986aj,Catani:1989ne,Catani:1991kz,Catani:1992ua,Korchemskaya:1992je,Korchemsky:1992xv,Korchemsky:1993uz,Forte:2002ni,Becher:2006nr,Becher:2007ty}
for a few seminal works in QCD and SCET. In
recent years, a great deal of work has been devoted
to understanding the factorization of physical
observables beyond leading power, and its
implications for the resummation of
large logarithms (see e.g.\
\cite{Bonocore:2016awd,Moult:2016fqy,DelDuca:2017twk,Beneke:2017ztn,Moult:2018jjd,Beneke:2018gvs,Moult:2019mog,Bahjat-Abbas:2019fqa,Moult:2019uhz,Beneke:2019oqx,Liu:2019oav,AH:2020iki,Engel:2021ccn,Czakon:2023tld}
for a selection of works, and references
therein). Resummation has been achieved
in general for leading logarithms (LLs)
at NLP. Phenomenological analyses
\cite{Beneke:2019mua,vanBeekveld:2021hhv}
have shown that the LLs at NLP are numerically
competitive with NNLLs at LP, see e.g.\
figures~\ref{Pheno-NLP-1} and~\ref{Pheno-NLP-2},
making their resummation highly relevant for
precision physics. Despite substantial progress
\cite{Beneke:2020ibj,Liu:2020tzd,Liu:2020wbn,Beneke:2022obx,Liu:2022ajh},
a systematic approach to NLP resummation
beyond LL accuracy is still missing.

\section{Factorization at leading power}

In what follows we continue to consider
Drell-Yan near threshold as a case study to
discuss the structure of NLP logarithms.
The simplicity of LP factorization
\cite{Sterman:1986aj,Catani:1989ne,Korchemsky:1993uz,
Forte:2002ni,Becher:2007ty}
\begin{equation}
\label{DYfact}
\frac{d\sigma}{dQ^2} =
|C^{A0}|^2 \times f_{a/A} \otimes f_{b/B}
\otimes S_{\rm DY} \big[Q(1-z)\big],
\end{equation}
originates from the fact that large
threshold logarithms are generated by
the emission of soft gluons only.
In the soft limit, a gluon with momentum
$k^\mu$ emitted from an external line with
momentum $p^\mu$ produces an eikonal factor,
\begin{equation}
\mathcal{M}\,\frac{\slashed{p}-\slashed{k}}{2p\cdot k}\,\gamma^\mu T^A u(p)
\sim \mathcal{M}\,\frac{p^\mu}{p\cdot k}\,T^A u(p),
\end{equation}
so that multi-gluon emission exponentiates
through Wilson lines:
\begin{equation}
\mathcal{M}\,\mathcal{S} \, u(p),
\quad {\rm with} \quad
\mathcal{S}=\langle 0|\Phi_\beta(-\infty,0)|0\rangle,
\end{equation}
and
$\Phi_\beta(\lambda_1,\lambda_2) = \mathcal{P} \exp \Big\{ig_s
\int_{\lambda_1}^{\lambda_2} d\lambda\,\beta\cdot A(\lambda\beta)\Big\}$.
The soft function in eq.~(\ref{DYfact}) thus encodes the factorization
and exponentiation of soft gluons.

\section{Factorization at next-to-leading power}

\begin{figure}[h]
\begin{center}
  \includegraphics[width=0.65\textwidth]{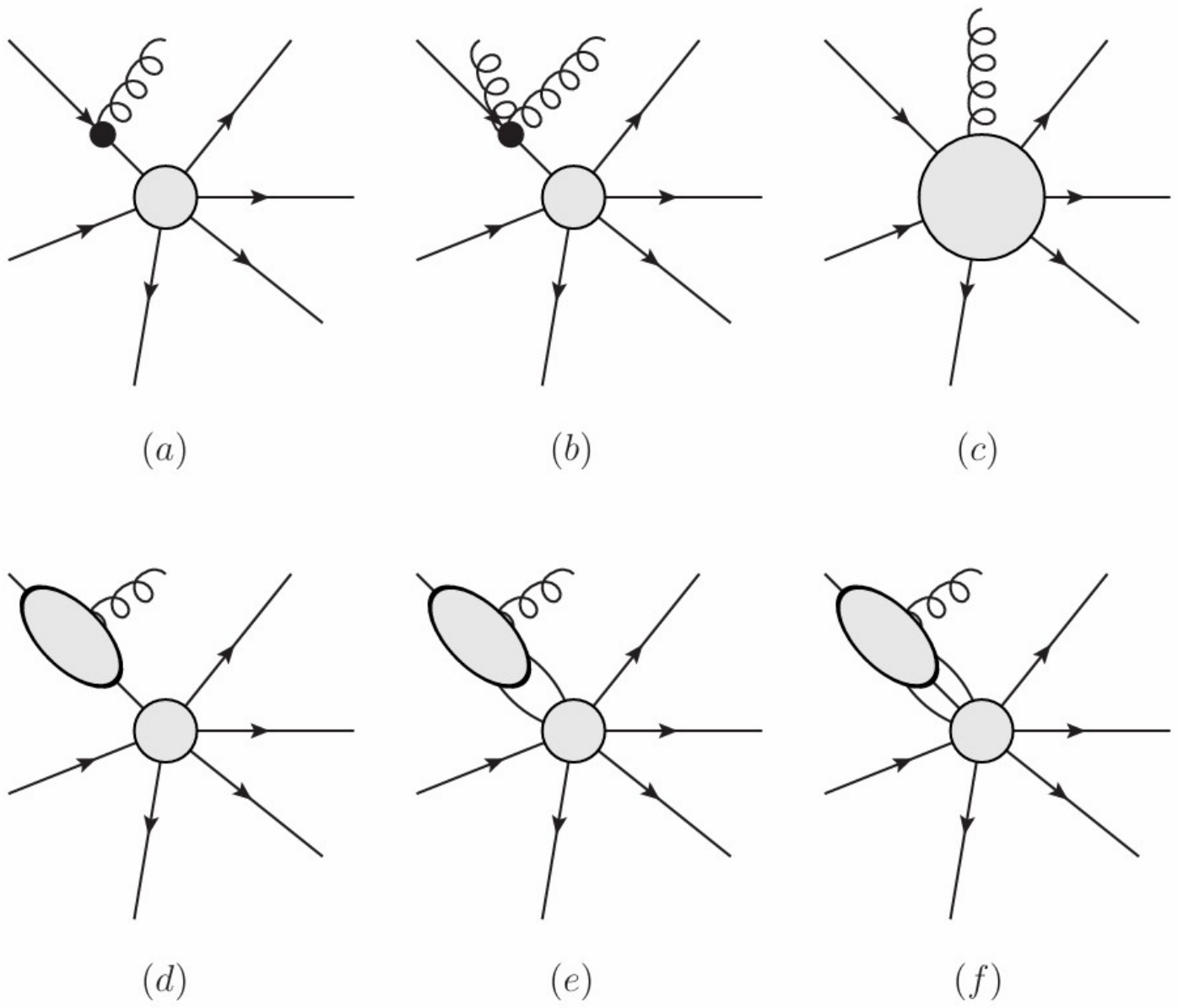}
\end{center}
\caption{Diagrammatic description of the contributions
appearing at NLP.}
  \label{NLP-structures}
\end{figure}
At NLP the factorization structure becomes significantly
richer. We distinguish three classes of power-suppressed
contributions appearing when factorizing soft and collinear
radiation from the hard interaction, either via SCET or via
a direct diagrammatic approach in QCD:
\begin{itemize}
\item \emph{Next-to-eikonal soft gluons}: emission beyond the
eikonal approximation, sensitive, for instance, to the spin of
the emitting particle (diagrams (a) and (b) in
figure~\ref{NLP-structures}), discussed in
a direct QCD approach in \cite{Laenen:2008gt,Laenen:2010uz}.
\item \emph{Direct emission from the hard interaction}:
this contribution (diagram (c) in figure~\ref{NLP-structures}),
first discussed by Low, Burnett and Kroll
\cite{Low:1958sn,Burnett:1967km}, accounts for the fact that
soft emissions begin to resolve the structure of the internal
hard non-radiative amplitude at next-to-leading order.
\item \emph{Radiative jets}: these describe the emission of soft gluons
from a cluster of virtual collinear particles in the same jet
sector (diagrams (d), (e) and (f) in figure~\ref{NLP-structures}),
first identified in \cite{DelDuca:1990gz} and later extended in
direct QCD in
\cite{Bonocore:2015esa,Bonocore:2016awd,Gervais:2017yxv,Laenen:2020nrt,vanBijleveld:2025ekz}.
Within SCET, these structures are identified with collinear
matrix elements \cite{Moult:2019mog,Beneke:2019oqx}.
\end{itemize}

\subsection{NLP factorization in SCET}

SCET \cite{Bauer:2000yr,Beneke:2002ph}
organizes the Lagrangian as $\mathcal{L}_{\rm SCET}
=\sum_i\mathcal{L}_{c_i}+\mathcal{L}_s + \sum_n\mathcal{O}_n $,
with operators
\begin{equation}
\mathcal{O}_n = \int\! dt_1\cdots dt_n\;
  \mathcal{C}(t_1,\ldots,t_n)\,\phi_1(t_1 n_{1+})\cdots\phi_n(t_n n_{n+}).
\end{equation}
The SCET Lagrangian and operators are constructed
so as to reproduce the momentum-region expansion
of a scattering cross section. Soft-collinear
decoupling at LP guarantees that the LP cross
section factorizes as
\begin{equation}
\sigma \sim \mathcal{H} \otimes \mathcal{J}_1
\otimes \cdots \otimes \mathcal{J}_n \otimes
\mathcal{S},
\end{equation}
where $\mathcal{H}$ is the hard matching coefficient,
$\mathcal{J}_i$ are jet functions (matrix elements of
collinear fields), and $\mathcal{S}$ is the soft
function. Renormalization group equations for each
function, which depends on a single scale, then resum
large logarithms.

At NLP one must account for subleading-power
Lagrangian insertions
$\mathcal{L}_{c_i}=\mathcal{L}_{c_i}^{(0)}
+\mathcal{L}_{c_i}^{(1)}+\cdots$, where
for instance \cite{Beneke:2002ph}
\begin{align}
\mathcal{L}_c^{(1)\,\rm gluon}(x) & =
  \bar{\xi}\!\left[x_\perp^\mu n_-^\nu W_c g_s F^s_{\mu\nu}(x_-)W_c^\dagger\right]
  \frac{\slashed{n}_+}{2}\xi, \\
\mathcal{L}_c^{(1)\,\rm quark}(x) & =
  \bar{q}(x_-)W_c^\dagger i\slashed{\partial}_{\perp c}\,\xi,
\end{align}
and power-suppressed operators such as
\begin{align}
J_\rho^{A0,A1}(t,\bar{t}) & =
\bar{\chi}_{\bar{c}}(\bar{t}n_-)n_{+\rho}
i\partial_\perp\chi_c(tn_+), \\
J_\rho^{A0,B1}(t_1,t_2,\bar{t}) & =
\bar{\chi}_{\bar{c}}(\bar{t}n_-)
n_{\pm\rho}\mathcal{A}_{\perp c}(t_2 n_+)\chi_c(t_1 n_+);
\end{align}
see \cite{Beneke:2017ztn,Beneke:2018rbh,Larkoski:2014bxa}
for further details.

A key complication is that factorization does not lead
straightforwardly to resummation: NLP convolutions in SCET
suffer from \emph{endpoint divergences}. For instance,
in the case of Drell-Yan, the factorization theorem involves
a convolution between the collinear and the soft function
of the form
\begin{equation}
\int_0^{\Omega} d\omega \,\underbrace{\big(n_+p
\, \omega\big)^{-\epsilon}}_{{\rm{collinear\,piece}}}
\,\underbrace{\frac{1}{\omega^{1+\epsilon}}
\frac{1}{(\Omega-\omega)^{\epsilon}}}_{{\rm{soft\,piece}}},
\end{equation}
which is divergent in $D=4$. As a consequence, standard RGE
methods must be supplemented by a non-trivial subtraction procedure
\cite{Beneke:2008pi,Liu:2019oav,Beneke:2022obx,Bell:2024bxg,Bell:2026ktc}.

\subsection{NLP factorization in direct QCD}

One of the aims of a direct QCD approach
is to determine the exponentiation
structure by means of a diagrammatic method,
thus avoiding (at least in the first instance)
the need to resort to renormalization-group
evolution and the problem of endpoint divergences.

In direct QCD several steps are analogous
to those of the SCET analysis,
with the difference that fields are not
divided into soft and collinear modes.
One starts by setting up a power counting,
obtained by decomposing momenta along the directions
of the external particles, $v^\mu = v^+ n_i^\mu
+ v^- \bar{n}_i^\mu + v_{\perp i}^\mu$, and
assigning scalings to all momentum components
and masses.
In general, the power counting coincides
with the scaling assigned by means of
expansion by regions \cite{Beneke:1997zp}.
For instance, for Drell-Yan near
threshold, collinear momenta scale as
$k^\mu\sim Q(1,\lambda,\lambda^2)$ with 
$\lambda \ll 1$, while soft momenta scale as
$k^\mu\sim Q(\lambda^2,\lambda^2,\lambda^2)$.
Next, using reduced diagrams (i.e.\ diagrams in 
which off-shell legs are contracted to a point, 
keeping on-shell lines), one determines the 
superficial degree of divergence as a function 
of the number of fermion and photon connections 
between the hard, soft, and collinear subgraphs
\cite{Sterman:1978bi,Collins:1989gx,Gervais:2017yxv,Laenen:2020nrt}.

\begin{table}[h]
\begin{center}
\begin{minipage}{0.48\linewidth}
\centering
\begin{tabular}{ll}
\hline
QED Vertex & Suppression \\ \hline
$\bar{\psi}^{(c)}\gamma^\mu\psi^{(c)}A_{\mu}^{(c)}$ & $\lambda$ \\
$\bar{\psi}^{(c)}\gamma^\mu\psi^{(c)}A_{\mu}^{(s)}$ & $1$ \\
$\bar{\psi}^{(s)}\gamma^\mu\psi^{(c)}A_{\mu}^{(c)}$,
$\bar{\psi}^{(c)}\gamma^\mu\psi^{(s)}A_{\mu}^{(c)}$ & $1$ \\
$\bar{\psi}^{(s)}\gamma^\mu\psi^{(s)}A_{\mu}^{(s)}$ & $1$ \\ \hline
\end{tabular}
\end{minipage}
\hfill
\begin{minipage}{0.48\linewidth}
\centering
\begin{tabular}{lll}
\hline
& $m=0$ & $m \sim \lambda Q$ \\ \hline
Collinear fermion & $\lambda^{-2}$ & \\
Soft fermion      & $\lambda^{-2}$ & $\lambda^{-1}$ \\
Collinear photon  & $\lambda^{-2}$ & \\
Soft photon       & $\lambda^{-4}$ & \\
Collinear loop    & $\lambda^{4}$  & \\
Soft loop         & $\lambda^{8}$  & \\ \hline
\end{tabular}
\end{minipage}
\end{center}
\caption{Left: power counting for QED vertices depending on the soft or collinear nature
of the fields (applies to massive and massless fermions alike).
Right: power counting for propagators and loop integrals; if no entry appears
for $m\sim\lambda Q$, the scaling is identical to the $m=0$ case.}
\label{tab:pcrules}
\end{table}
This determines all diagrammatic structures
contributing to a given scattering amplitude
up to the desired power in the expansion.
For instance, in QED, with momentum and mass
scalings as in table~\ref{tab:pcrules},
the complete list of NLP collinear contributions
to the amplitude reads
\cite{Laenen:2020nrt}
\begin{align}
\label{factLP}
\mathcal{M}^{\rm LP} &= \bigg(\prod_{i=1}^n J_{(f)}(\hat{p}_i)\bigg)
  \otimes H(\hat{p}_1,\ldots,\hat{p}_n)\,S(n_i\cdot n_j), \\ \nonumber
\label{factNLP}
\mathcal{M}_{\rm coll}^{\rm NLP} &=
+ \sum_{i=1}^n \bigg(\prod_{j\neq i}J_{f}^j\bigg)
\Big[J^i_{f\gamma} \otimes H^i_{f\gamma}
+J^i_{f\partial \gamma} \otimes H^i_{f\partial \gamma}\Big]\,S \\ \nonumber
&\quad + \sum_{i=1}^n \bigg(\prod_{j\neq i}J_{f}^j\bigg)
J^i_{f\gamma\gamma} \otimes H^i_{f\gamma\gamma}\, S
+ \sum_{i=1}^n \bigg(\prod_{j\neq i}J_{f}^j\bigg)
J^i_{f\!f\!f} \otimes H^i_{f\!f\!f}\, S  \\
&\quad +\sum_{1\leq i \leq j \leq n}
\bigg(\prod_{k\neq i,j}J_{f}^k \bigg)
J^i_{f\gamma}J^j_{f\gamma}
\otimes H^{ij}_{f\gamma,f\gamma}\, S
+ {\cal O}(\lambda^3)\,.
\end{align}
In the first line the momentum
dependence is written explicitly, with $\hat{p}_i^\mu=p_i^+n_i^\mu$
representing the large momentum component of a collinear
momentum along direction $i$. Contributions at NLP
involve multiple particles in the same collinear sector,
which induces a convolution over the momentum
fraction along the common collinear direction
(subleading collinear limits). For instance, the
first term in $\mathcal{M}_{\rm coll}^{\rm NLP}$
reads more explicitly:
\begin{align} \label{eq:explicitFirstLine} \nonumber
\bigg(\prod_{j\neq i}J_{(f)}^j\bigg)
\Big[J^i_{(f\gamma)} \otimes H^i_{(f\gamma)}
+J^i_{(f\partial \gamma)} \otimes H^i_{(f\partial \gamma)}\Big]\,S
&\equiv
S(\hat{p}_i\cdot\hat{p}_j;\epsilon)\,
\bigg(\prod_{j\neq i}J_{(f)}(p_{j};\epsilon)\bigg)  \\ \nonumber
&\hspace{-5.0cm}\times\, \int_0^{p_i^+}d \ell^+_i
\bigg[J^\nu_{(f\gamma)}(p_i-\hat{\ell_i},\hat{\ell_i};\epsilon)\,
H_{(f\gamma)\nu}(p_1 \dots;p_i-\hat{\ell_i},\hat{\ell_i};\dots p_n;\epsilon)\, \\
&\hspace{-3.5cm}+\,J^{\nu\rho}_{(f\partial\gamma)}(p_i-\hat{\ell_i},\hat{\ell_i};\epsilon)\,
H_{(f\partial\gamma)\nu\rho}(p_1 \dots;p_i-\hat{\ell_i},\hat{\ell_i};\dots p_n;\epsilon)\,
\bigg]\,.
\end{align}

The next step consists in finding gauge-invariant
operator matrix element definitions for the jet and
soft functions in QED/QCD, such that the power-suppressed
collinear and soft configurations are correctly reproduced
\cite{vanBijleveld:2025ekz}. For instance, for the LP jet one
finds
\begin{equation}\label{JfDef}
J_f(p_1,\bar{n}) = \langle p_1|\bar{\psi}(0)\,\Phi_{\bar{n}}(0,\infty)|0\rangle,
\end{equation}
whilst the NLP ``radiative'' jet carrying an additional
photon with momentum fraction $\bar{n}\cdot\ell$ in QED
reads
\begin{equation}\label{JfgammaDef}
J_{f\gamma}^\rho(p_1,\bar{n},\bar{n}\cdot\ell)
= \int_{-\infty}^{\infty}\!\frac{d\xi}{2\pi}\,e^{-i\ell(\xi\bar{n})}
  \langle p_1\big|
  \big[\bar{\psi}(0)\Phi_{\bar{n}}(0,\infty)\big]
  \big[\Phi_{\bar{n}}(\infty,\xi\bar{n})
       \bigl(iD^\rho\Phi_{\bar{n}}(\xi\bar{n},\infty)\bigr)\big]
  \big|0\rangle.
\end{equation}
The jet and soft functions
are defined so as to reproduce
the collinear and soft
region results obtained by means of
a momentum-region expansion, respectively.
A key difference with respect to the
operators introduced in SCET is that
in SCET each operator has a definite
power counting, and the SCET Lagrangian
is constructed such that collinear
and soft matrix elements give rise
(in particular in the case of SCET-I)
exclusively to collinear and soft
contributions. The price to pay is
relatively involved Feynman rules
associated with the subleading-power
Lagrangian and operators. By contrast, 
each of the functions introduced above 
is itself organised as a series in the 
power expansion, with its own well-defined 
leading order (LP for $J_f$, N$^{1/2}$LP for
$J_{f\gamma}^\rho$, NLP for the others), 
followed by the full tower of 
subleading-power terms. Furthermore, the jet
functions may also contain
soft contributions that need
to be removed by a subtraction
procedure. On the other hand,
one of the advantages of the
functions in eq.~(\ref{factNLP})
is that they can be calculated
directly in QCD.

\section{\boldmath
A case study: the massive quark form factor
for $m^2 \ll s$}

The massive electromagnetic form factor with
$m^2 \ll s$ provides the simplest non-trivial
check of the framework, being the first case
in which all NLP jet functions in eq.~(\ref{factNLP})
beyond $J_{f}$ are required \cite{terHoeve:2023ehm,vanBijleveld:2025ekz}
to reproduce the vertex function, defined
as follows:
\begin{equation}\label{ampDef}
V^\mu(p_1,p_2) = \bar{v}(p_2)\,\Gamma^\mu(p_1,p_2)\,u(p_1),
\end{equation}
where
\begin{equation}
\Gamma^\mu = -ie\,e_q\!\left[F_1(s,m^2)\gamma^\mu
  +\frac{1}{2m}F_2(s,m^2)\,i\sigma^{\mu\nu}q_\nu\right].
\end{equation}
Calculating the form factors in the limit
$m^2 \ll s$ by means of the method of
regions is already quite useful for
singling out factorization properties beyond
leading power. Among the various possible
regions,
\begin{equation}
\begin{array}{cc}
&
k^{\mu} = (k^+, k^-, k_{\perp}) \\
\text{hard ($h$):} \qquad
&
k\sim\sqrt{\hat{s}}\,
(\lambda^0,\lambda^0,\lambda^0)\,, \\
\text{collinear ($c$):} \qquad
&
k\sim\sqrt{\hat{s}}\,
(\lambda^0,\lambda^2,\lambda^1)\,, \\
\text{anti-collinear ($\bar{c}$):} \qquad
&
k\sim\sqrt{\hat{s}}\,
(\lambda^2,\lambda^0,\lambda^1)\,, \\
\text{ultra-collinear ($uc$):} \qquad
&
k\sim\sqrt{\hat{s}}\,
(\lambda^2,\lambda^4,\lambda^3)\,, \\
\text{ultra-anti-collinear ($\overline{uc}$):} \qquad
&
k\sim\sqrt{\hat{s}}\,
(\lambda^4,\lambda^2,\lambda^3)\,, \\
\text{semi-hard ($sh$):} \qquad
&
k\sim\sqrt{\hat{s}}\,
(\lambda^1,\lambda^1,\lambda^1)\,, \\
\text{soft ($s$):} \qquad
&
k\sim\sqrt{\hat{s}}\,
(\lambda^2,\lambda^2,\lambda^2)\,,
\end{array}
\end{equation}
the semi-hard and soft regions
contribute only in the presence of
soft radiation, which is not the case
for the form factors. At one loop, the hard,
collinear, and anti-collinear regions
contribute: for instance, for the
form factor $F_1$ one finds
\begin{align} \nonumber
F_1^{(1l)}\Big|_h = &
\bigg(\frac{\mu^2}{-\hat s -i0^+}\bigg)^{\epsilon}
\bigg\{ - \frac{2}{\epsilon^2} - \frac{3}{\epsilon}
- 8 + \zeta_2 + \epsilon \bigg(-16 + \frac{3\zeta_2}{2}
+ \frac{14\zeta_3}{3}\bigg) \\ \nonumber
&\hspace{2.0cm}
+\, \frac{m^2}{\hat s} \bigg[- \frac{2}{\epsilon} -6
+ \epsilon \big(-16 + \zeta_2\big) \bigg]
+{\cal O}(\epsilon^2)+ {\cal O}(\lambda^4) \bigg\}, \\ \nonumber
F_1^{(1l)}\Big|_c =& \bigg(\frac{\mu^2}{m^2-i0^+}\bigg)^{\epsilon}
\bigg\{\frac{1}{\epsilon^2} + \frac{2}{\epsilon} + 4
+ \frac{\zeta_2}{2} + \epsilon \bigg(8 + \zeta_2
- \frac{\zeta_3}{3}\bigg) \\
&\hspace{2.0cm}
+\, \frac{m^2}{\hat s} \bigg[\frac{1}{\epsilon}
+ 5 + \epsilon \bigg(13 + \frac{\zeta_2}{2}\bigg) \bigg]
+{\cal O}(\epsilon^2)+ {\cal O}(\lambda^4) \bigg\}.
\end{align}

\begin{figure}
    \centering
    \includegraphics[width=.24\textwidth]{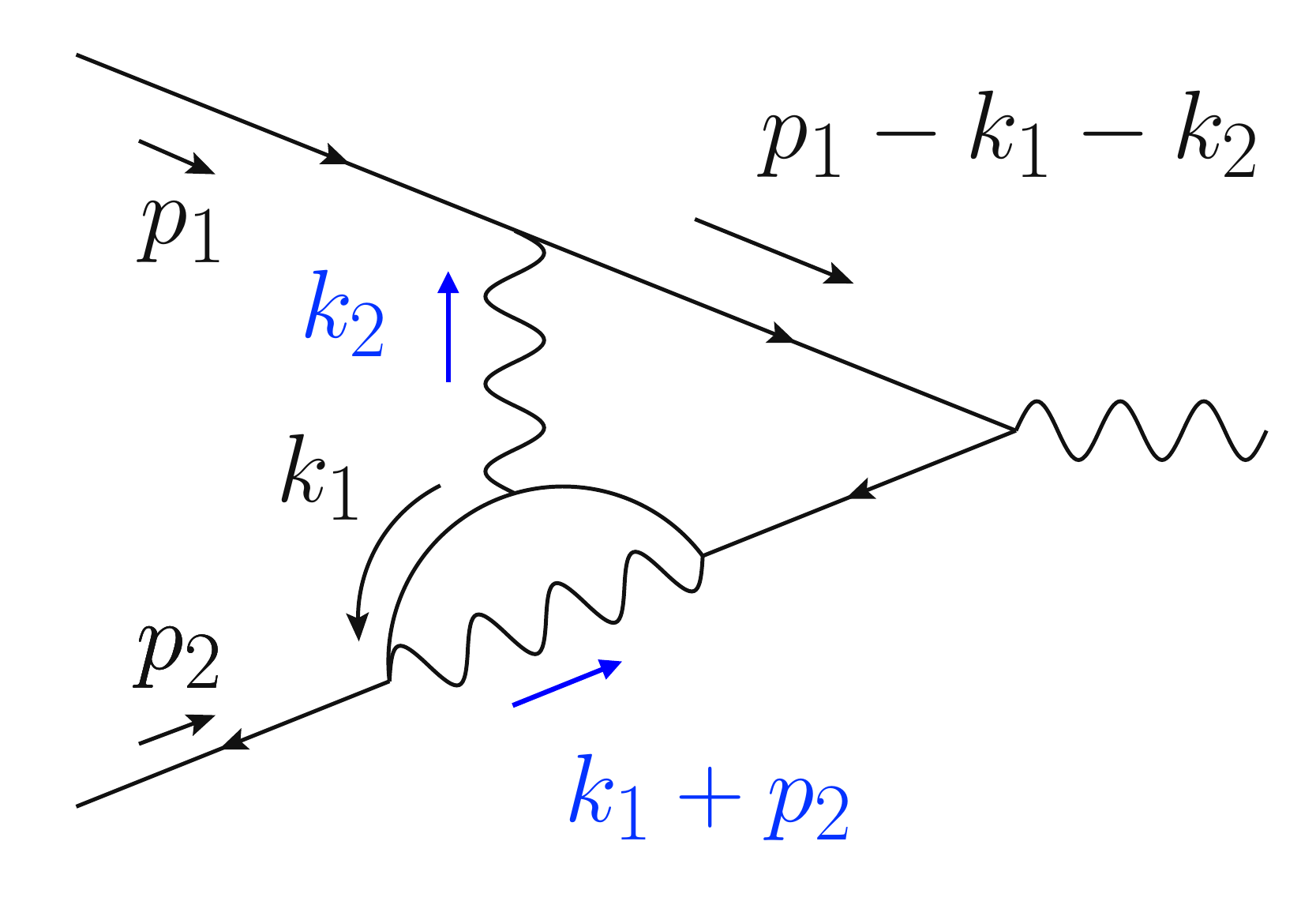}
    \,\,\,
    \includegraphics[width=.72\textwidth]{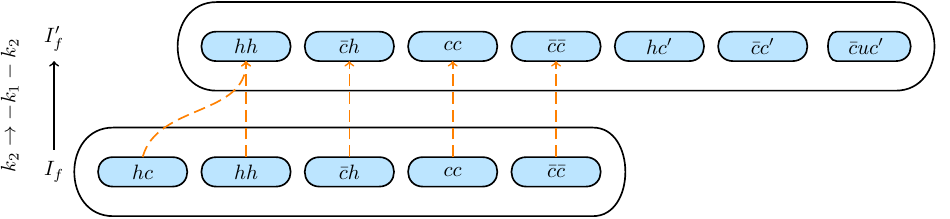}
    \caption{Momentum regions contributing to the scalar integral associated with the
    diagram on the left-hand side, before and after applying the transformation
    $k_2 \rightarrow -k_1 -k_2$, corresponding to $I_f$ and $I_f'$ respectively. The
    dashed arrows indicate how the regions in $I_f$ transform accordingly; e.g.\ the
    original $hc$-region maps onto a new $hh$-region after the collinear mode $k_2$
    mixes with the dominant hard scale associated with $k_1$. Note how two previously
    absent regions, $\bar{c}c'$ and $\bar{c}uc'$, and a different $hc$-region
    appear after the shift, where the $\bar{c}c'$-region cancels the rapidity
    divergences present in the $cc$ and $\bar{c}\bar{c}$ regions. Regions that
    remain invariant are displayed on top of each other, whilst additional regions
    are shifted outwards, so that all regions are found by collapsing the top
    row onto its base $I_f$.}
    \label{fig:regions_shift_diag_f}
\end{figure}
The two-loop calculation
(see \cite{Bernreuther:2004ih,Gluza:2009yy}
for the original references with the exact
result in QCD) is more involved, but also
more revealing. Let us begin by noting that
all regions can be identified only by carefully
choosing the momentum routing: for instance,
in the case of the scalar master integral
associated with the diagram in
figure~\ref{fig:regions_shift_diag_f}, all
regions can be singled out only by
considering both loop-momentum
parametrizations, as indicated on
the right-hand side of the same figure.
A crucial observation is that
additional collinear regions
(ultra-collinear and ultra-anti-collinear)
appear at the level of individual diagrams but
cancel in the contribution to the
form factors. This fact, already
observed at LP in \cite{Becher:2007cu},
confirmed here at NLP, and proven recently 
to all orders at LP in \cite{Schnubel:2026gee}), 
provides an important consistency check 
for the factorization programme: if new 
regions were to contribute at each subsequent 
loop order, any attempt to establish a 
factorization theorem would be invalidated.

Once the two-loop form factors have been
calculated within the method of regions
and checked against the literature, one needs
to calculate the jet functions according
to their matrix element definitions; see
e.g.\ eqs.~(\ref{JfDef}) and~(\ref{JfgammaDef}).
For instance, for $J_f$ and $J_{f\gamma}$
at ${\cal O}(\alpha_s)$ one finds
\cite{vanBijleveld:2025ekz}
\begin{align}\nonumber
J_{f}^{(1)}(p_1,\bar{n}) & =
i 16\pi^2\bar{u}(p_1)\int[dk]\,
\frac{\slashed{\bar{n}}
(\slashed{p}_1-\slashed{k}+m)}{k^2
[(p_1-k)^2-m^2][\bar{n}\cdot k]} \\[0.2cm]
& = \left(\frac{\bar\mu^2}{m^2}\right)^\epsilon
\bar{u}(p_1)\, \frac{\Gamma(\epsilon)
e^{\epsilon\gamma_E}}{\epsilon(1-2\epsilon)}
\bigg(1-\frac{m\epsilon}{p_1^+} \slashed{\bar{n}}\bigg), \\[0.1cm] \nonumber
J_{f\gamma}^{(1)\rho}(p_1,\bar{n},\ell^+)
& = i 16\pi^2\, \bar{u}(p_1)\int[dk]
\left(\eta^{\rho\sigma}
-\frac{\bar{n}^\sigma k^\rho}{\bar{n}\cdot k}\right)
\frac{\gamma_\sigma\left(\slashed{p}_1-\slashed{k}
+m\right)}{k^2[(k-p_1)^2-m^2]}
\delta(\bar{n}\cdot k -\ell^+) \\[0.2cm] \nonumber
& = \left(\frac{\bar\mu^2}{m^2}\right)^{\epsilon}
\Gamma(\epsilon)e^{\epsilon\gamma_E}\, \bar{u}(p_1)
\, m \, \Bigg\{x^{1-2\epsilon}
\left(-\gamma^\rho+\frac{\slashed{\bar{n}}\,
\hat{p}_1^\rho}{p_1^+}\right) \\
&\hspace{1.0cm}+\,\frac{m}{p_1^+} \bigg[\frac{1}{2(1-\epsilon)}
\left(\delta(1-x)-(1-2\epsilon)x^{1-2\epsilon}\right)
\gamma^\rho \slashed{\bar{n}}-2x^{-2\epsilon}(1-x)
\bar{n}^\rho\bigg] \Bigg\}\,.
\end{align}
These results are checked by specialising
eq.~(\ref{factNLP}) to the $q\bar{q}$ amplitude
in eq.~(\ref{ampDef}) and expanding up to two
loops:
\begin{align} \nonumber
V^{\mu(1)} &= J_f^{(0)}\,H_{f,\bar{f}}^{\mu(1)}\,J_{\bar{f}}^{(0)}
  + J_f^{(1)}\,H_{f,\bar{f}}^{\mu(0)}\,J_{\bar{f}}^{(0)}
  + J_f^{(0)}\,H_{f,\bar{f}}^{\mu(0)}\,J_{\bar{f}}^{(1)} \\ \nonumber
&+\,  J_{f\gamma}^{(1)}\otimes H_{f\gamma,\bar{f}}^{\mu(0)}\,J_{\bar{f}}^{(0)}
  + J_f^{(0)}\,H_{f,\bar{f}\gamma}^{\mu(0)}\otimes J_{\bar{f}\gamma}^{(1)}, \\[0.2cm]
V^{\mu(2)} &= J_f^{(0)}\,H_{f,\bar{f}}^{\mu(2)}\,J_{\bar{f}}^{(0)}
  + J_f^{(2)}\,H_{f,\bar{f}}^{\mu(0)}\,J_{\bar{f}}^{(0)}
  + J_f^{(0)}\,H_{f,\bar{f}}^{\mu(0)}\,J_{\bar{f}}^{(2)}
  + J_f^{(1)}\,H_{f,\bar{f}}^{\mu(1)}\,J_{\bar{f}}^{(0)}+\cdots,
\end{align}
where in the two-loop case only the first few terms are displayed.
Comparing with the result
obtained with the method of regions, one has
for instance at one loop:
\begin{equation}
V^{\mu (1)}_{f,\bar f}
= V^{\mu (1)}_{f,\bar f}\Big|_h
+ V^{\mu (1)}_{f,\bar f}\Big|_c
+ V^{\mu (1)}_{f,\bar f}\Big|_{\bar c},
\end{equation}
so that one expects
\begin{align} \nonumber
V^{\mu (1)}_{f,\bar f}\Big|_h &= J_f^{(0)}\,H_{f,\bar{f}}^{\mu(1)}\,J_{\bar{f}}^{(0)}, \\ \nonumber
V^{\mu (1)}_{f,\bar f}\Big|_c &= J_f^{(1)}\,H_{f,\bar{f}}^{\mu(0)}\,J_{\bar{f}}^{(0)}
+ J_{f\gamma}^{(1)}\otimes H_{f\gamma,\bar{f}}^{\mu(0)}\,J_{\bar{f}}^{(0)}, \\
V^{\mu (1)}_{f,\bar f}\Big|_{\bar c} &= J_f^{(0)}\,H_{f,\bar{f}}^{\mu(0)}\,J_{\bar{f}}^{(1)}
+ J_f^{(0)}\,H_{f,\bar{f}\gamma}^{\mu(0)}\otimes J_{\bar{f}\gamma}^{(1)}.
\end{align}
At two loops one proceeds similarly, and we refer to
\cite{vanBijleveld:2025ekz} for further details. We
note that the two-loop result involves all types of
jet function appearing in eq.~(\ref{factNLP}), so
the factorization formula is fully verified by
our case study.

\section{Conclusions}

In these proceedings we have reviewed the current
state of the art in the factorization and resummation
of hadronic cross sections near threshold at
next-to-leading power (NLP). Two complementary
frameworks have been discussed: Soft-Collinear
Effective Theory (SCET) and a direct diagrammatic
approach in QCD.

Within SCET, a factorization theorem at NLP is
well established. The cross section factorizes into
hard, collinear (jet), and soft functions, each
carrying a single characteristic scale. However,
proceeding from factorization to resummation meets
a fundamental obstacle: the convolutions between
collinear and soft functions suffer from endpoint
divergences, which prevent a straightforward
application of renormalization-group methods.
At leading-logarithmic accuracy these divergences
can be handled by means of subtraction procedures
\cite{Beneke:2008pi,Liu:2019oav,Beneke:2022obx,Bell:2024bxg,Bell:2026ktc},
and resummation of LLs at NLP has been achieved
in this way for several processes
\cite{Beneke:2018gvs,Beneke:2019mua,vanBeekveld:2021hhv}.
Beyond LL, however, a systematic treatment of
endpoint divergences is still missing, and
constitutes the main open problem on the SCET side.

In direct QCD, substantial progress has been made
towards a complete factorization theorem at NLP.
After setting up a power counting via momentum-region
techniques, all diagrammatic configurations
contributing at NLP can be classified, and
gauge-invariant operator matrix element definitions
for the jet and soft functions can be provided.
As a non-trivial check of the framework, we have
discussed the massive quark form factor in the
limit $m^2\ll s$, for which all NLP jet functions
entering the factorization formula~(\ref{factNLP})
are required. The factorization formula has been
verified explicitly at one and two loops
\cite{terHoeve:2023ehm,vanBijleveld:2025ekz},
confirming that only hard and (anti-)collinear
regions contribute through NLP, and that
ultra-collinear regions present in individual
diagrams cancel in the physical amplitude.

To extend the factorization theorem to physical
cross sections, one must go beyond the virtual
amplitude and include real radiation, thereby making
the jet functions \emph{radiative}. The
definitions of radiative jet functions used
so far in the literature
\cite{DelDuca:1990gz,Bonocore:2016awd} are
based on the insertion of a conserved current,
which by construction describes the emission
of a single soft photon or gluon. For
resummation and exponentiation, however,
one needs to describe the emission of an
arbitrary number of soft gluons. Understanding
the structure of radiative jets with multiple
soft-gluon emissions therefore represents
a key open problem for the NLP programme
in direct QCD.

Interestingly, the situation is more
favourable on the purely soft side.
At NLP, both soft gluons and soft quarks
can be described within a unified framework
based on \emph{generalised soft functions}
and \emph{generalised Wilson lines}.
The generalised soft function
$\mathcal{S}$ can be written as a path
integral over generalised Wilson lines,
which encode eikonal, next-to-eikonal
spin-independent, spin-dependent, and
seagull contributions. Applying the
replica trick \cite{Gardi:2010rn},
one can show that $\mathcal{S}$
exponentiates in terms of \emph{generalised webs}
\cite{Laenen:2010uz,vanBeekveld:2021mxn,vanBeekveld:2023liw}:
\begin{align} \nonumber
{\cal \widetilde S}
= \langle 0 |F_1 \ldots F_n| 0\rangle
& \sim e^{\sum_{G_e} \, C_{G_e} {\cal W}_{G_e}
+\sum_{G_{ne}} \, C_{G_{ne}} {\cal W}_{G_{ne}}} \\
&\sim e^{\sum_{G_e} \, C_{G_e} {\cal W}_{G_e}}
\Big( 1+\sum_{G_{ne}} \, C_{G_{ne}} {\cal W}_{G_{ne}} \Big),
\end{align}
where $F_n$ represent generalised Wilson
lines, and ${\cal W}_{G_e}$, ${\cal W}_{G_{ne}}$
are respectively leading- and next-to-leading power
(eikonal and next-to-eikonal) webs,
provided the phase space for $n$-gluon
emission factorizes into $n$ decoupled
single-gluon integrals, a condition
satisfied at LP and at LL NLP
\cite{vanBeekveld:2023liw}.
The exponentiation of the purely
soft sector at NLP is thus, in
principle, a solved problem.
Given this result, understanding
the exponentiation of radiative jets
with an arbitrary number of emitted
gluons constitutes the last
missing ingredient for a complete
understanding of the exponentiation
structure of QCD cross sections
near threshold at NLP.

\end{document}